\documentclass[useAMS,usenatbib]{mn2e}

\usepackage{times}
\usepackage{graphicx}

\newcommand{\beq}{\begin{equation}}
\newcommand{\eeq}{\end{equation}}

\def\gs{\mathrel{\lower0.6ex\hbox{$\buildrel {\textstyle >}\over{\scriptstyle \sim}$}}}
\def\ls{\mathrel{\lower0.6ex\hbox{$\buildrel {\textstyle <}\over{\scriptstyle \sim}$}}}

\begin{document}

\title[Deprojecting galaxy clusters]{On the deprojection of clusters of galaxies combining X-ray, Sunyaev-Zeldovich temperature decrement and gravitational lensing maps}
\author[M. Sereno]{M. Sereno\thanks{E-mail:sereno@physik.unizh.ch} 
\\
Institut f\"{u}r Theoretische Physik, Universit\"{a}t Z\"{u}rich,
Winterthurerstrasse 190, CH-8057 Z\"{u}rich , Switzerland
}


\maketitle

\begin{abstract}
Knowledge of the intrinsic shape of galaxy clusters is very important in investigating cosmic structure formation and astrophysical processes. The reconstruction of the 3-dimensional structure usually relies on deprojecting 2-dimensional X-ray, Sunyaev-Zeldovich (SZ) and/or gravitational lensing observations. As known, a joint analysis of these data sets can provide the elongation of the cluster along the line of sight together with its length and width in the plane of the sky. An unbiased measurement of the Hubble constant can be also inferred. Due to some intrinsic degeneracies, the observational constraints obtained from such projected data-sets are not enough to allow an unique inversion. In general, the projected maps can be at the same time compatible with prolate, oblate and with many triaxial configurations. Even a prolate cluster might be interpreted as an oblate system and vice versa. Assuming that the cluster is axially symmetric is likely to overestimate the intrinsic ellipticity, whereas the system always looks rounder performing the inversion under the hypothesis of a triaxial cluster aligned with the line of sight. In general, analysing triaxial clusters under the prolate or oblate assumption may introduce strong biases even when the clusters are actually near to axial symmetry whereas the systematics introduced assuming the cluster to be aligned with the line of sight are more under control.
\end{abstract}


\section{Introduction}

The determination of the intrinsic shape of astronomical objects is a classic topic. The intrinsic structure of galaxies or cluster of galaxies directly probes the cosmic structure formation suggesting how material aggregates from large-scale perturbations \citep{wes94}. It also contains evidence about the nature and mechanisms of interaction of baryons and dark matter, since processes such as virialization, dissipation or gas cooling tend to make systems more spherical, especially in the inner regions \citep{kaz+al04}. 

The complex structure of halos also affects mass estimates \citep{gav05} and could cause a significant bias in estimating the inner density matter slope and the concentration parameter of matter halos \citep{ogu+al05}. These quantities are crucial for any attempt at high precision cosmology and when comparing observations with theoretical predictions from numerical simulations \citep{voi05}. 

The first attempts to determine three dimensional morphologies were based on statistical approaches consisting in the inversion of the distribution of apparent shapes. \citet{hub26} first determined the relative frequencies with which galaxies of a given intrinsic ellipticity, oriented at random, are observed as having various apparent projected ellipticities. Several following studies have then applied similar methods to different classes of astronomical objects~\citep{noe79,bin80,bi+de81,fa+vi91,det+al95,moh+al95,bas+al00,coo00,th+ch01,al+ry02,ryd96,pli+al04,paz+al06}. With the exception of disc galaxies, prolate-like shapes appear to dominate all cosmic structure on a large scale.

Together with statistical studies, the deprojection of single objects was also investigated. Based on the Fourier slice theorem, \citet{ryb87} argued  that, due to a cone of ignorance in the Fourier space, the deprojection can not be unique, even assuming axial symmetry. \citet{ge+bi96} further showed that either discy or boxy 3-D distributions can be compatible with the same projected image.

Clusters of galaxies have the strong advantage that their structure can be routinely probed with very heterogeneous data-sets at very different wave-lengths. This consideration boosted studies on how combining X-ray surface brightness and spectral observations of the intra-cluster medium (ICM), Sunyaev-Zeldovich effect (SZE) and gravitational lensing (GL) observations. One of the main result that can be obtained combining different data-sets is breaking the degeneracy between the line of sight elongation and the distance to the cluster, which allows in principle an unbiased estimate of the Hubble constant \citep{fo+pe02}. The deprojection of the density distribution was also reconsidered. \citet{zar+al98} proposed a method based on the axial symmetry assumption and on the extrapolation of the image Fourier transform into the cone of ignorance. Alternative procedures can be based either on the iterative Richardson Lucy deconvolution and again assuming axial symmetry \citep{reb00,pu+ba06} or on a perturbation approach \citep{dor+al01}. \citet{def+al05} and \citet{ser+al06} finally applied a joint X-ray plus SZE parametric analysis to a sample of $25$ clusters, finding that prolate rather than oblate shapes seem to be preferred, with signs of a more general triaxial morphology.

Many of the above methods, despite very insightful, often rely on very restrictive assumptions, sometimes not clearly stated, and aim to obtain very general results. The methods are then usually tested with the application to some numerical simulations, so that possible degeneracies in the deprojection techniques can be easily over-sought. In this paper I propose a simple analytical discussion aimed to explore what are the features of the cluster shape that we can really measure by combining X-ray, SZE and gravitational lensing data and under which conditions the deprojection can be performed unequivocally. The paper is as follows. In Sec.~\ref{sec_proj}, I describe how a 3-D ellipsoid casts on the plane of the sky. Section~\ref{sec_mult} briefly presents the main characteristics of X-ray, SZE and GL observations, whereas Sec.~\ref{sec_join} discusses which constraints about cluster shape and orientation can be inferred. In Sec.~\ref{sec_part}, I consider when only a single deprojection is allowed, whereas Sec.~\ref{sec_appr} treats the error made on the determination of the cluster shape under particular assumptions. Section~\ref{sec_badl} is devoted to when and how a prolate cluster can appear as oblate and vice-versa. At the end, Section~\ref{sec_disc} contains some final considerations.

\section{Projected ellipsoids}
\label{sec_proj}

High resolution $N$-body simulations have clearly shown that the density profiles of matter halos are aspherical and how such profiles can be accurately described by concentric triaxial ellipsoids with aligned axes \citep{ji+su02}. Then, the electron density of the intra-cluster medium can be assumed to be constant on a family of similar, concentric, coaxial ellipsoids. The intra-cluster medium (ICM) distribution in clusters of galaxies in hydrostatic equilibrium traces the gravitational potential. Since we are considering a triaxial elliptical gas distribution, the gravitational potential turns out to be constant on a family of similar, concentric, coaxial ellipsoids. Elliptical gravitational potential can turn unphysical for extreme axial ratios, giving negative density regions or very unlikely configurations, but as far as inner regions are considered, they can provide very suitable approximations. 

In an intrinsic orthogonal coordinate system centred on the cluster's barycentre and whose coordinates are aligned with its principal axes, a spheroidal ICM profile can be described by only one radial variable
$\zeta$,
\beq
\zeta^2 \equiv \sum_{i=1}^3 e_i^2 x_{i,\mathrm{int}}^2 .
\label{eq:rad_var}
\eeq
Along each axis, $e_i$ is the inverse of the corresponding core radius in units of a scale-length $r_{\rm c}$. Without loss of generality, we can fix $e_3=1$, which means that our reference scale-length is the core radius along $x_\mathrm{3,int}$. For an axially symmetric cluster, if the polar axis is aligned with the third coordinate axis, $e_i= \left\{ 1/q_\mathrm{int}, 1/ q_\mathrm{int}, 1 \right\}$ for a prolate model and $e_i= \left\{q_\mathrm{int}, q_\mathrm{int}, 1 \right\}$ for an oblate model, where $q_\mathrm{int} \le 1$ is the intrinsic axial ratio.

When viewed from an arbitrary direction, quantities constant on similar ellipsoids project themselves on similar ellipses~\citep{sta77}. Three rotation angles relate the intrinsic to the observer's coordinate system, i.e. the three Euler's angles, $\theta_{\rm Eu}, \varphi_{\rm Eu}$ and $\psi_{\rm Eu}$ of the three principal cluster axes with respect to the observer. A rotation through the first two Euler's angles is sufficient to align the $x_{3,\rm obs}$-axis of the observer coordinates system $\left\{x_{i,\rm obs} \right\}$ with whatever direction. In what follows, I will assume that the $x_{3,\rm obs}$-axis is aligned with the line of sight to the observer, i.e. the direction connecting the observer to the cluster centre. Then, in the intrinsic system, the line of sight has polar angles $\{ \theta, \phi \}=\{ \theta_{\rm Eu},\varphi_{\rm Eu} -\pi/2 \}$. With a third rotation, $\psi_{\rm Eu}$, we can properly align the $x_{1,\rm obs}$- and $x_{2,\rm obs}$-axis in the plane of the sky. In general, $\psi_{\rm Eu}$ is the angle in the plane of the sky between the projection of the $x_\mathrm{3,int}$-axis and the $x_\mathrm{2,obs}$-axis. Unfortunately, the direction of the $x_\mathrm{3,int}$-axis is not known, so that we can not choose a reference system in the plane of the sky such that $\psi_{\rm Eu} = 0$. Then, if not stated otherwise we will line up the $x_{1,\rm obs}$- and $x_{2,\rm obs}$-axis with the axes of the projected ellipses.

The ellipticity and the orientation of the projected ellipses depend only on the intrinsic geometry and orientation of the system. The axial ratio of the major to the minor axis of the observed projected isophotes, $e_{\rm p}(\geq 1)$, can be written as \citep{bin80},
\begin{equation}
\label{eq:tri4e}
e_{\rm p}= \sqrt{ \frac{j+l + \sqrt{(j-l)^2+4 k^2 } }{j+l
-\sqrt{(j-l)^2+4 k^2 }} },
\end{equation}
where  $j, k$ and $l$ are defined as
\begin{eqnarray}
j & = &  e_1^2 e_2^2 \sin^2 \theta_{\rm Eu} + e_1^2 \cos^2 \theta_{\rm Eu} \cos^2 \varphi_{\rm Eu} \nonumber \\
  & + &  e_2^2 \cos^2 \theta_{\rm Eu} \sin^2 \varphi_{\rm Eu} ,  \label{eq:tri4a} \\
k & = &  (e_1^2 - e_2^2) \sin \varphi_{\rm Eu} \cos \varphi_{\rm Eu} \cos \theta_{\rm Eu}  ,  \label{eq:tri4b}  \\
l & = &  e_1^2 \sin^2 \varphi_{\rm Eu} + e_2^2 \cos^2 \varphi_{\rm Eu} . \label{eq:tri4c}
\end{eqnarray}

As written before, the projected direction of the $x_{3,\rm int}$-axis is not known. If we assume that the coordinate axes in the plane of the sky lie along the axes of the isophotes then \citep{bin85}
\begin{equation}
\label{eq:tri4f}
\psi_\mathrm{Eu} = \frac{1}{2} \arctan \left[\frac{2 k}{j-l} \right].
\end{equation}
The apparent principal axis that lies furthest from the projection of the $x_{3,\rm int}$-axis onto the plane of the sky is the apparent
major axis if \citep{bin85}
\begin{equation}
(j-l)\cos 2 \psi_\mathrm{Eu} +2k \sin 2\psi_\mathrm{Eu} \leq 0
\end{equation}
or the apparent minor axis otherwise. In general the right-hand side of Eq.~(\ref{eq:tri4f}) is the angle between the principal axes of the observed ellipses and the projection onto the sky of the $x_\mathrm{3,int}$-axis.

The observed cluster angular core radius $\theta_\mathrm{p}$ is the projection on the plane of the sky of the cluster angular intrinsic core radius \citep{sta77},
\begin{equation}
\label{eq:tri6}
\theta_{\rm p} \equiv \theta_{\rm c} \left( \frac{e_{\rm p}}{e_1 e_2} \right)^{1/2} f^{1/4}
\end{equation}
where $\theta_{\rm c} \equiv r_{\rm c}/D_{\rm d}$, with $D_\mathrm{d}$ the angular diameter distance to the cluster, and $f$ is a function of the
cluster shape and orientation,
\begin{equation}
\label{eq:tri3}
f = e_1^2 \sin^2 \theta_{\rm Eu} \sin^2 \varphi_{\rm Eu} + e_2^2 \sin^2 \theta_{\rm Eu} \cos^2 \varphi_{\rm Eu} + \cos^2 \theta_{\rm Eu} .
\end{equation}

\section{Multi-Wavelength Observations}
\label{sec_mult}

In this section, I briefly summarise the main features of X-ray, SZE and GL observations. As a reference model, we consider a cluster electron density distribution, $n_\mathrm{e}$, described by an ellipsoidal triaxial $\beta$-model. In the intrinsic coordinate system,
\begin{equation}
\label{eq:tri0}
n_\mathrm{e}  =  n_\mathrm{e0} \left( 1+ \frac{\zeta^2}{r_{\rm c}^2} \right)^{-3\beta/2}
\end{equation}
where $\beta$ is the slope, $r_{\rm c}$ is the core radius and $n_\mathrm{e0}$ is the central electron density. We are interested in observed quantities which are given by projection along the line of sight of powers of the electron density $n_\mathrm{e}$. The projection of a generic power $m$ of $n_\mathrm{e}$, as given in Eq.~(\ref{eq:tri0}), can be written as \citep{def+al05}
\begin{eqnarray}
\label{eq:tri5}
\int _{\rm l.o.s.} n_\mathrm{e}^m ( l )  dl &  = &  n_\mathrm{e0}^m \sqrt{\pi}
\frac{\Gamma \left[3 m\beta/2-1/2 \right]}{\Gamma \left[3 m\beta/2
\right]} \frac{D_{\rm d} \theta_{\rm c}}{\sqrt{f}} \nonumber \\
& \times & \left( 1+ \frac{\theta_{1}^2+e_{\rm p}^2 \theta_{2}^2}{\theta_\mathrm{p}^2} \right)^{(1-3 m\beta)/2}
\end{eqnarray}
where $\theta_i \equiv x_{i,\rm obs}/D_{\rm d}$ is the projected angular position on the plane of the sky of $x_{i,\rm obs}$. The quantity $r_\mathrm{c}/\sqrt{f}$ in Eq.~(\ref{eq:tri5}) represents the half-size of the ellipsoid along the line of sight. It can be conveniently rewritten as
\beq
\label{mult1}
\frac{r_\mathrm{c}}{\sqrt{f}} \equiv \frac{r_\mathrm{p}}{e_\Delta},
\eeq 
which represents the definition of the elongation $e_\Delta$; $r_\mathrm{p}(\equiv D_\mathrm{d}\theta_\mathrm{p})$ is the projected radius.If $e_\Delta < 1$, then the cluster is more elongated along the line of sight than wide in the plane of the sky. In terms of the elongation, the projected core radius can be expressed as
\begin{equation}
\theta_{\rm p} \equiv \theta_{\rm c} \left( \frac{e_{\rm p} e_\Delta}{e_1 e_2} \right)^{1/3}.
\end{equation}

In a Friedmann-Lema\^{\i}tre-Robertson-Walker universe filled with pressure-less matter and with a cosmological
constant, the angular diameter distance between the redshift $z_{\rm d}$ and a source at $z_{\rm s}$ is \citep[and references therein]{ser+al01}
\begin{equation}
\label{eq:crit3}
D (z_{\rm d}, z_{\rm s})=\frac{c}{H_0}\frac{1}{1+z_{\rm
s}}\frac{1}{|\Omega_{\rm K0}|} {\rm Sinn} \left( \int_{z_{\rm
d}}^{z_{\rm s}} \frac{|\Omega_{\rm K0}|}{{\cal E}(z)} dz \right)
\end{equation}
with
\begin{eqnarray}
\label{eq:dist1}
{\cal E}(z) & \equiv & \frac{H(z)}{H_0} \\ & =& \sqrt{ \Omega_{\rm M0}
(1+z)^3+ \Omega_{\Lambda 0} + \Omega_{\rm K0}(1+z)^2}
\nonumber
\end{eqnarray}
where $H_0$, $\Omega_{\rm M0}$ and $\Omega_{\Lambda 0}$ are the Hubble parameter, the normalised energy density of pressure-less matter and
the reduced cosmological constant at $z=0$, respectively. $\Omega_{\rm K0}$ is given by $\Omega_{\rm K0}\equiv 1- \Omega_\mathrm{M0}-\Omega_{\Lambda 0}$, and Sinn is defined as being $\sinh$ when $\Omega_{\rm K0}>0$, $\sin$ when $\Omega_{\rm K0}<0$, and as the identity when $\Omega_{\rm K0}=0$.

\subsection{X-ray Surface Brightness}

Cluster X-ray emission is due to bremsstrahlung and line radiation resulting from electron-ion collisions in the high temperature plasma ($k_{\rm B} T_{\rm e} \approx 8$-$10\ {\rm keV}$, with $k_{\rm B}$ being the Boltzmann constant). The X-ray surface brightness $S_X$ can be written as
\begin{equation}
S_X = \frac{1}{4 \pi (1+z )^4} \int _{\rm l.o.s.} n_\mathrm{e}^2 \Lambda_\mathrm{e}(T_\mathrm{e}, {\cal{Z}}) dl ,
\label{eq:sxb0}
\end{equation}
where $\Lambda_\mathrm{e}$ is the cooling function of the ICM in the cluster rest frame and depends on the ICM temperature $T_\mathrm{e}$ and metallicity ${\cal{Z}}$. Assuming an isothermal plasma with constant metallicity and taking the result from Eq.~(\ref{eq:tri5}) for $m=2$, we get
\begin{equation}
S_X = S_{X0} \left( 1+ \frac{\theta_{1}^2+e_{\rm p}^2
\theta_{2}^2}{\theta_\mathrm{p}^2} \right)^{1/2-3 \beta}  ,
\label{eq:sxb1}
\end{equation}
where the central surface brightness $S_{X0}$ reads
\begin{equation}
\label{eq:sxb2}
S_{X0} \equiv \frac{ \Lambda_\mathrm{e} }{4 \sqrt{\pi} (1+z)^4}
n_\mathrm{e0}^2  \frac{D_{\rm d} \theta_{\rm p}} {e_\Delta} \frac{ \Gamma (3\beta -1/2)}{\Gamma (3\beta)} .
\end{equation}

\subsection{The Sunyaev-Zeldovich Effect}

Photons of the cosmic microwave background (CMB) that pass through the hot ICM of a cluster interact with its energetic electrons through inverse Compton scattering, slightly distorting the CMB spectrum. This is the Sunyaev-Zeldovich effect (SZE)~\citep{su+ze70,bir99}, which is proportional to the electron pressure integrated along the line of sight. The measured temperature decrement $\Delta T_{\rm SZ}$ of the CMB is given by,
\begin{equation}
\label{eq:sze1}
\frac{\Delta T_{\rm SZ}}{T_{\rm CMB}} = f_{SZ}(\nu, T_{\rm e}) \frac{ \sigma_{\rm T} k_{\rm B} }{m_{\rm e} c^2}
\int _{\rm l.o.s.}n_\mathrm{e} T_{\rm e} dl ,
\end{equation}
where $T_{\rm CMB}$ is the temperature of the CMB, $\sigma_{\rm T}$ the Thompson cross section, $m_{\rm e}$ the electron mass, $c$ the speed of light in vacuum and $f_{SZ}(\nu, T_{\rm e})$ accounts for relativistic corrections at frequency $\nu$. For an isothermal $\beta$-model, taking the result in Eq.~(\ref{eq:tri5}) for $m=1$, we obtain,
\begin{equation}
\label{eq:sz2}
\Delta T_{\rm SZ} = \Delta T_0 \left( 1+ \frac{\theta_{1}^2+e_{\rm p}^2 \theta_{2}^2}{\theta_\mathrm{p}^2}
\right)^{1/2-3\beta/2} ,
\end{equation}
where $\Delta T_0$ is the central temperature decrement which includes all the physical constants and the terms resulting from the line of sight integration
\begin{eqnarray}
\label{eq:sze3}
\Delta T_0 & \equiv & T_{\rm CMB} f_{SZ}(\nu, T_{\rm e}) \frac{ \sigma_{\rm T} k_{\rm B} T_{\rm e}}{m_{\rm e} c^2} n_\mathrm{e0} \sqrt{\pi}\nonumber \\
&\times&  \frac{D_{\rm d}\theta_\mathrm{p}}{e_\Delta}  \frac{ \Gamma (3\beta/2 -1/2)}{\Gamma (3\beta/2)} .
\end{eqnarray}

\subsection{Gravitational Lensing}
\label{sec:lens}

Clusters of galaxies act as lenses deflecting light rays from background galaxies. In contrast to SZE and X-ray emission,
gravitational lensing does not probe the ICM but maps the total mass. The cluster total mass can be related to its gas distribution if the intra-cluster gas is assumed to be in hydrostatic equilibrium in the cluster gravitational potential. If we assume that the gas is isothermal and that non-thermal processes do not to contribute significantly to the gas pressure, the total dynamical mass density can be expressed as
\begin{equation}
\label{eq:gl1}
\rho_{\rm tot} = -\left( \frac{k_\mathrm{B} T_{\rm e}}{4 \pi G \mu m_{\rm p}} \right)
\nabla^2 \left( \ln n_{\rm e} \right) ,
\end{equation}
where $G$ is the gravitational constant and $\mu m_{\rm p}$ is the mean particle mass of the gas. Ellipsoidal ICM distributions determine ellipsoidal potentials, which are widely used in gravitational lensing analyses \citep{sef}. 

If we assume that the ICM follows a $\beta$-model distribution, the projected mass density can be subsequently derived \citep{def+al05}.  The lensing effect is determined by the convergence $k=\Sigma/ \Sigma_{\rm cr}$ which is the cluster surface mass density in units of the critical density $\Sigma_{\rm cr}$,
\begin{equation}
\label{eq:gl5}
\Sigma_{\rm cr} \equiv \frac{c^2}{4 \pi G} \frac{D_{\rm s}}{D_{\rm d} D_{\rm ds}} ,
\end{equation}
where $D_{\rm ds}$ is the angular diameter distance from the lens to the source and $D_{\rm s}$ is the angular diameter distances from the observer to the lens. The convergence reads \citep{def+al05}
\begin{equation}
\label{eq:gl3}
k = k_0 \left(1+ \frac{e_{\rm p}^2}{1+e_{\rm p}^2}
\frac{\theta_{1}^2+ \theta_{2}^2}{\theta_{\rm p}^2} \right)
\left( 1+ \frac{\theta_{1}^2+e_{\rm p}^2 \theta_{2}^2}{\theta_\mathrm{p}^2} \right)^{-3/2}
\end{equation}
where
\begin{equation}
k_0=3 \pi \beta \frac{ k_{\rm B} T_{\rm e}}{c^2 \mu m_{\rm p}}
\frac{ 1 }{e_\Delta} (1+e_{\rm p}^2) \frac{1}{\theta_\mathrm{p}}
\frac{D_{\rm ds}}{D_{\rm s}} .
\label{eq:gl6b}
\end{equation}
The convergence can be measured through either a weak lensing analysis of statistical distortion of images of background galaxies or by fitting the observed surface mass density to multiple image strong lensing systems. Although the hypotheses of hydrostatic equilibrium and isothermal gas are very strong, total mass densities obtained under such assumptions can yield accurate estimates even in dynamically active clusters with irregular X-ray morphologies \citep{def+al04}.

\section{Joint analysis of heterogeneous data sets}
\label{sec_join}

This section discusses which geometrical constraints on the cluster shape can be inferred from projected maps.

\subsection{Unknown parameters}

The intrinsic shape of an ellipsoidal cluster is described by two axis ratios, $e_1$ and $e_2$. Its orientation is fixed by three Euler's angles, $\theta_{\rm Eu}, \phi_{\rm Eu}$ and $\psi_{\rm Eu}$. 

The density ICM profile is characterised by some other scale parameters. For a $\beta$-model, three additional parameters come in: the central density normalisation, $n_\mathrm{e0}$, the slope index, $\beta$, and the core radius $r_{\rm c}$. Then, $5+3$ parameters characterise the ICM distribution. Under the hypothesis of isothermality, a single value, $T_{\rm e}$, characterises the temperature of the cluster. If the metallicity is nearly constant, then one other parameter $\cal{Z}$ is enough to describe this quantity. 

The cosmological dependence enters through the cosmological distances. For a flat model of universe, the distance-redshift relation is determined by two parameters: the Hubble constant, $H_0$, and the matter density parameter $\Omega_{\rm M0}$.

\subsection{Observational constraints}

As seen before, 3-D ellipsoids are casted in 2-D projected ellipses. By fitting an elliptical profile to the X-ray and/or SZE data, both the projected axis ratio, $e_{\rm p}$, and the orientation angle can be measured providing two constraints on the intrinsic shape 
\begin{eqnarray}
e_\mathrm{p}       & = & e_\mathrm{p} (e_1, e_2; \theta_\mathrm{Eu}, \vartheta_\mathrm{Eu}) , \label{con1}\\
\psi_\mathrm{Eu} & = & \psi_\mathrm{Eu} (e_1, e_2; \theta_\mathrm{Eu}, \vartheta_\mathrm{Eu}) .\label{con2}
\end{eqnarray}
Eq.~(\ref{con2}) expresses the freedom to align the isophotes with the coordinate axis.

Together with the shape and orientation of the ellipses, the fitting procedure can also provide further constraints on the density profile. In particular, for a $\beta$-model, the slope $\beta$ and the projected core radius $\theta_\mathrm{p}$ can also be determined from data.
\begin{eqnarray}
\beta_\mathrm{obs} & = & \beta, \label{con3}  \\
\theta_\mathrm{p}         & = & \theta_\mathrm{p} (e_1, e_2; e_\Delta(e_1, e_2; \theta_\mathrm{Eu}, \vartheta_\mathrm{Eu}) ), \label{con4}
\end{eqnarray}
where we have made clear the dependence on the unknown quantities.

Besides these constraints on the density profile, the temperature of the ICM, as well as its metallicity, can be inferred from spectroscopic X-ray observations with sufficient spectral resolution
\begin{eqnarray}
T_\mathrm{e}    & = & T_\mathrm{obs} , \label{con5}    \\ 
{\cal{Z}}& = & {\cal{Z}}_\mathrm{obs}. \label{con5a}
\end{eqnarray}

The observed values of the central surface brightness, $S_{X0}$, Eq.~(\ref{eq:sxb2}), of the central temperature decrement, $\Delta T_0$, Eq.~(\ref{eq:sze3}), and of the gravitational lensing convergence, Eq.~(\ref{eq:gl6b}), provide three further constraints, 
\begin{eqnarray}
S_{X0}       & \propto & \frac{ n_\mathrm{e0}^2}{e_\Delta} D_\mathrm{d}          , \label{con6}  \\
\Delta T_{0} & \propto & \frac{ n_\mathrm{e0}}{e_\Delta} D_\mathrm{d}            , \label{con7} \\
k_{0}        & \propto & \frac{1}{e_\Delta} \frac{D_\mathrm{ds}}{D_\mathrm{s}}   . \label{con8}
\end{eqnarray}
The dependence on the elongation of the cluster shows up both in the relation between the projected and intrinsic core radius, Eq.~(\ref{con4}), and in the expression of the central quantities, Eqs.~(\ref{con6}-\ref{con8}). The convergence $k$ depends on the cosmology through the ratio of distances $D_{\rm ds}/D_{\rm s}$. Therefore $k$ depends only the cosmological density parameters $\Omega_i$, and not on the Hubble constant $H_0$. In all, we have two observational constraints less than the unknown parameters, so that the system is under-constrained.

\subsection{Inferred quantities}

Let us see what we can learn on the cluster structure. Equation~(\ref{con4}) and Eqs.~(\ref{con5},~\ref{con5a}) refers specifically to an isothermal $\beta$-model with constant metallicity. In any case, they show a general feature: the parameters which describe the temperature and density profile can be derived in principle with accurate spectroscopic and photometric observations even for more complicated models. Instead, we are mainly interested in the intrinsic shape (2 parameters) and orientation (3 parameters) of the cluster. As a first step, let us assume that the cosmological density parameters are independently known, so that cosmological distances are known apart from an overall factor proportional to the Hubble constant, $D_i \propto c/H_{0}$. Then, the system of Eqs.~(\ref{con6}-\ref{con8}) is closed and can be simply solved,
\begin{eqnarray}
n_\mathrm{e0}    & \propto & \frac{ S_{X0}}{\Delta T_{0}}     ,  \label{con9} \\
e_\Delta  & \propto & \frac{1}{k_0}                      ,       \label{con11} \\
H_0       & \propto & k_0 \frac{S_{X0}}{\Delta T_{0}^2}  .       \label{con10}
\end{eqnarray}
As well known, combining SZ and X-ray observations only fixes the central gas density. On the other hand, the degeneracy between the Hubble constant and the physical size of the cluster along the line of sight can be broken only with the additional information provided by gravitational lensing under the assumption of hydrostatic equilibrium \citep{fo+pe02}.

Observations of either multiple strong lensing image systems or weak lensing of background sources with well determined photometric redshifts could provide further information on the cosmological parameters $\Omega_i$. In fact, the value of $k_0$ changes according to the redshift $z_\mathrm{s}$ of the lensed source through the ratio $D_{\rm ds}/{D_{\rm s}}$, so that for each background source redshift we have an additional constraint on the geometry of the universe. With a sufficient number of image systems at different redshifts an estimate of all cosmological parameters involved could therefore be performed \citep{ser02}. Stacking gravitational lensing data of systems at different redshifts will then provide information on the cosmology and will reduce the observational uncertainties on the density parameters but will not add constraints on the intrinsic shape of the cluster. One image system is enough to break the degeneracy between the Hubble constant and the elongation of the cluster. On the other hand, if we trust independent estimates of the Hubble constant, then SZ and X-ray data are enough to measure the elongation of the cluster and we do not need the additional theoretical constraint of hydrostatic equilibrium and the observational gravitational lensing data.

From the above analysis it is clear that even relying on extraordinarily accurate projected X-ray, SZE and GL maps, the intrinsic shape of the cluster can not be unambiguously inferred. The only information that we can establish on the cluster is its width and length in the plane of the sky, i.e. $r_\mathrm{p}$ and $r_\mathrm{p}/e_\mathrm{p}$, and its size along the line of sight, $r_\mathrm{p}/e_\Delta$; $r_\mathrm{p}$  is related to the intrinsic scale-length through Eq.~(\ref{con4}). Furthermore, we have a relation which expresses the third Euler's angle, $\psi_\mathrm{Eu}$ in terms of the other intrinsic parameters. These 4 constraints are what an astronomer dealing with projected X-ray, SZE and GL maps can use to determine the (2) intrinsic axial ratios, the (3) orientation angles and the (1) intrinsic length-scale of the cluster.

\subsection{On a general density distribution}

The isothermal $\beta$ model can be sometimes inaccurate \citep{ras+al06} but the conclusions on what we can learn on the 3-D structure of galaxy clusters from projected maps do not change considering more accurate models. The temperature profile should account for a central cool region, if any, and a gradient al large radii \citep{vik+al06}. In the same way, the ICM density profile should be able to account for a possible central steep increase of the surface brightness and a change of slope at larger radii \citep{vik+al06}. Accurate modelling is required when studying the ICM physics or when reliable mass estimates have to be obtained from deep observations but, as far as an insight on which physical constraints on the 3-D shape we can get from projected maps is concerned, the conclusions are not affected. Constraints obtained using more accurate models are more reliable but do not allow to break any degeneracy discussed in the framework of the $\beta$ model. In fact degeneracies are connected to the ellipsoidal intrinsic structure, not to the specific features of the radial profile. Then, whatever the model used, what can be inferred on the cluster intrinsic structure is its elongation, ellipticity and orientation of the isophotes, together with an estimate of the Hubble constant (if GL data are available) and with the parameters characterising the properties of the density distribution (slope, concentration,...). 

Under very broad assumptions, i.e. ICM density profile monotonically decreasing with radius, any radial distribution (temperature, metallicity,...) can be expressed as a function of the ICM distribution (or of the gravitational potential). Furthermore, with the hypothesis of hydrostatic equilibrium, the temperature profile is described by the same axis ratios and orientation of the ICM. Then, both the X-ray surface brightness, Eq.~(\ref{eq:sxb1}), and the SZ temperature decrement, Eq.~(\ref{eq:sze1}), are projections of some function of $n_\mathrm{e}$. Furthermore, the temperature profile can be in principle extracted taking ratios of the X-ray emission in different energy bands and is therefore independent of the elongation \citep{fo+pe02}. Then, a more detailed temperature analysis would constrain important features of the temperature profile, such as a possible cool core radius, but it would not help in providing additional independent constraints on the shape and orientation of the cluster.

In general, an intrinsic volume density $F_\mathrm{v}$ and its projection on the plane of the sky $F_\mathrm{s}$ are related to by \citep{sta77}
\beq
\label{mult2}
F_\mathrm{s} (\xi) = \frac{2}{\sqrt{f}} \int_\xi^\infty F_\mathrm{v} \frac{\zeta}{\sqrt{ \zeta^2-\xi^2}} d \zeta
\eeq
where $\xi$ is the ellipsoidal radius in the plane of the sky. For coordinate axes oriented along the isophotes, $\xi^2 = D_\mathrm{d}^2 (\theta_1^2 + e_\mathrm{p}^2 \theta_2^2) f/e_\Delta$; $\xi^2$ can be rewritten in terms of generic intrinsic ($l_\mathrm{c}$) and projected ($l_\mathrm{p}$) scale-lengths, related as in Eq.~(\ref{mult1}), i.e. $\xi^2 = D_\mathrm{d}^2 (\theta_1^2 + e_\mathrm{p}^2 \theta_2^2) (l_\mathrm{c}/l_\mathrm{p})^2$. Since the integral in $\zeta$ in Eq.~(\ref{mult2}) is proportional to the intrinsic scale length, we can write
\beq
\label{mult3}
F_\mathrm{s} \propto D_\mathrm{d}\frac{\theta_\mathrm{p}}{e_\Delta} f_\mathrm{s} (e_\mathrm{p}, \psi_\mathrm{Eu}; \theta_\mathrm{p}; p_i,...) 
\eeq
where as usual $l_\mathrm{p} = D_\mathrm{d}\theta_\mathrm{p}$ and $p_i$ are the parameters describing the 3-D density function $F_\mathrm{v}$. The structure of Eq.~(\ref{mult3}) is the same of Eqs.~(\ref{con6}) and (\ref{con7}), which were obtained for a specific model. The dependence on the elongation $e_\Delta$ is decoupled from the dependence on the apparent ellipticity and inclination and the parameters characterising the 3-D profile only accounts for the radial dependence of the projected density. Then, as $\beta$ can be derived by fitting the projected isophotes to a $\beta$-model, the parameters of a different profile (slope, concentration, ...) can be determined by a similar procedure as well. As before, the intrinsic scale-lengths are related to their projected values as in Eq.~(\ref{mult1}). So the statement that the only information that we can establish on the cluster is its sizes in the plane of the sky and along the line of sight does not depend on the particular mass model of the cluster.

The above considerations show that luminosity and surface brightness observations in the optical band (which are other example of projected maps) would provide the same kind of information on the intrinsic structure as those obtained by X-ray and SZE observations.

\section{Particular solutions}
\label{sec_part}

\begin{figure*}
        \resizebox{\hsize}{!}{\includegraphics{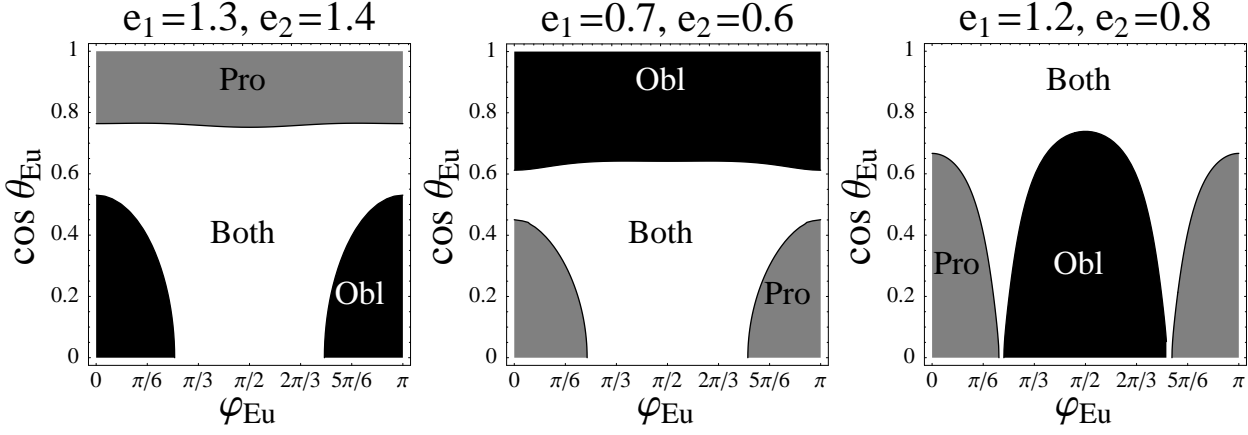}}
        \caption{Solution domains as a function of the orientation angles of the line of sight for three different triaxial ellipsoids. The grey, black and white regions denote the loci of line of sight directions in which only the prolate, only the oblate and both solutions are admissible, respectively. $e_3$ is fixed to $1$. Left panel: a nearly prolate ellipsoid, $e_1 = 1.3$, $e_2=1.4$.  Middle panel: a nearly oblate ellipsoid, $e_1 = 0.7$, $e_2=0.6$.  Right panel: a pretty triaxial ellipsoid, $e_1 = 1.2$, $e_2=0.8$.
}
	\label{Fig_ProOrObl}
\end{figure*}

\begin{figure}
        \resizebox{\hsize}{!}{\includegraphics{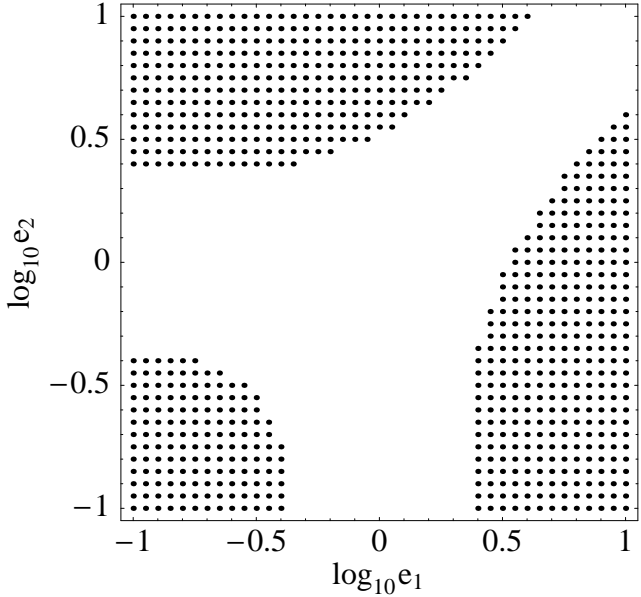}}
        \caption{Solution domains as a function of the intrinsic axis ratio. $e_3$ is fixed to $1$. The white and black spotted regions denote the loci of axial ratio for which the double and the oblate solution are more likely, respectively.
}
	\label{Fig_ProOrObl_e1e2}
\end{figure}

Having ascertained that a full inversion can not be performed only based on 2-D maps, I now turn on the possibility of deprojecting cluster observations under particular assumptions. Let us first consider some intrinsic degeneracies. As well known, when dealing with projected maps, we are not able to determine which extremity of the cluster is pointing towards the observer, $e_i (\theta_\mathrm{Eu}) = e_i (-\theta_\mathrm{Eu})$, $i=\{p, \Delta\}$. That is why in what follows we will limit our considerations to $0 \le \theta_\mathrm{Eu} \le \pi/2$. In the same way, the observed quantities are invariant for $\varphi_\mathrm{Eu} \rightarrow \varphi_\mathrm{Eu} + \pi$. Elongation and projected ellipticity are also symmetric around $\varphi_\mathrm{Eu} = \pi/2$, i.e. $e_\Delta (\pi/2 + \varphi_\mathrm{Eu}) = e_\Delta (\pi/2 - \varphi_\mathrm{Eu})$ and the same for $e_\mathrm{p}$. Furthermore for $\varphi_\mathrm{Eu} = \pi/4$, $e_\mathrm{p}$ and $e_\Delta$ do not change for $e_1 \rightarrow e_2$ and $e_2 \rightarrow e_1$.

One simple way to break intrinsic degeneracies is to fix the orientation of the line of sight in the intrinsic system of the ellipsoid. This lowers the number of unknown parameters by bringing out $\theta_\mathrm{Eu}$ and $\varphi_\mathrm{Eu}$. Since observations can determine the elongation and $e_\mathrm{p}$ and since the third Euler's angle can be expressed as function of the axial ratios and the other two angles, one could think that assuming that $\theta_\mathrm{Eu}$ and $\varphi_\mathrm{Eu}$ are known, then the inversion to determine $e_1$ and $e_2$ could be unambiguously performed. Unfortunately, the equations for $e_1$ and $e_2$
\begin{eqnarray}
e_\Delta (e_1, e_2; \left. \theta_\mathrm{Eu} \right|_\mathrm{fix}, \left. \varphi_\mathrm{Eu} \right|_\mathrm{fix}) & = &  \left.  e_\Delta \right|_\mathrm{obs}, \\
e_\mathrm{p} (e_1, e_2; \left. \theta_\mathrm{Eu} \right|_\mathrm{fix}, \left. \varphi_\mathrm{Eu} \right|_\mathrm{fix}) & = &  \left.  e_\mathrm{p} \right|_\mathrm{obs},
\end{eqnarray}
are not linear. Multiple solutions can exist even when the orientation of the system is correctly known. The situation gets even worse when the orientation is not exactly fixed to the actual direction. When performing a deprojection, one can make an attempt to fix the inclination angles to some particular (trial) values and then check if the inversion is possible under this assumption. Unfortunately, choosing a trial orientation for the line of sight we are not even assured about the existence of even one solution. Just as an example, if the line of sight of a triaxial system with $e_1 = 0.7$ and $e_2 =1.2$ lies along $\cos \theta_\mathrm{Eu} =1/2$ and $\varphi_\mathrm{Eu}=\pi/3$, then the equation system for $e_1$ and $e_2$ does not have a real solution even if the trial values are fixed to $\left. \cos \theta_\mathrm{Eu} \right|_\mathrm{trial} = 1/2 $  and $(\left. \varphi_\mathrm{Eu} \right|_\mathrm{trial}- \varphi_\mathrm{Eu})/\varphi_\mathrm{Eu} =10^{-2}$, i.e with a very small error $\Delta \varphi_\mathrm{Eu} \sim 0.5\deg$. On the other hand, for $(\left. \varphi_\mathrm{Eu} \right|_\mathrm{trial}- \varphi_\mathrm{Eu})/\varphi_\mathrm{Eu} =- 10^{-2}$, there are two solutions.

It can then be useful to consider some particular assumptions under which the solution exists and is unique. Interesting configurations are either an aligned triaxial or an axially symmetric ellipsoid. Under the assumption that one of the principal axis of the ellipsoid is aligned with the line of sight, the inversion can be easily performed and the intrinsic axial ratios are easily recovered. This hypothesis was exploited in \citet{def+al05} to deproject a sample of luminous X-ray clusters with SZE observations. As an example, when the $x_\mathrm{3,int}$-axis is aligned with the line of sight, i.e. $\theta_\mathrm{Eu} =0 $ and $\varphi_\mathrm{Eu}=\pi/2$, then the maximum between $e_1$ and $e_2$ will be $e_\mathrm{p}/e_\Delta$ and the minimum $1/e_\Delta$ (as usual $e_3=1$).

Another very popular choice is assuming that the cluster shape is nearly axially symmetric \citep{ser+al06}. In this case the ICM distribution is characterised by just two parameters: the ratio of the minor to the major axis, $q_\mathrm{int}(\le 1)$, and the inclination angle, $i$, between the line of sight and the polar axis. If the polar axis lies along the $x_\mathrm{3,int}$-axis, then $i = \theta_\mathrm{Eu}$. The major (minor) axis of the isophotes coincides with the projection of the polar angle if the ellipsoid is prolate (oblate). A useful parameter to quantify the triaxiality degree of an ellipsoid is
\beq
T=\frac{e_\mathrm{mid} - e_\mathrm{min}}{e_\mathrm{max} - e_\mathrm{min} },
\eeq
where $e_\mathrm{min}$, $e_\mathrm{mid}$ and $e_\mathrm{max}$ are $e_1$, $e_2$ and $e_3$ sorted in growing order. Oblate and prolate clusters correspond to $T=0$ and $1$, respectively.

The projected ellipticity and the elongation  of a prolate cluster can be easily expressed in terms of the intrinsic parameters \citep{ser+al06},
\begin{eqnarray}
 e_\mathrm{p}       & = &  \frac{\sqrt{1- \left(1-q_{\mathrm{int}}^2\right)\cos ^2 i}}{q_{\mathrm{int}}},  \label{pro1}\\
 e_\Delta  & = &  \frac{1- \left(1-q_{\mathrm{int}}^2\right)\cos ^2 i}{q_{\mathrm{int}}}  .\label{pro2}
\end{eqnarray}
The previous couple of equations can be then easily inverted to infer the intrinsic shape,
\begin{eqnarray}
 q_\mathrm{int}        & = &  \frac{e_{\Delta }}{e_\mathrm{p}^2}  ,\label{pro3}   \\
 \cos i    & = &  e_\mathrm{p} \sqrt{\frac{e_\mathrm{p}^2-e_{\Delta }^2}{e_\mathrm{p}^4-e_{\Delta }^2}}  .  \label{pro4}
\end{eqnarray}
A prolate-like solution is then admissible only when the size along the line of sight is larger then the minimum width in the plane of the sky, i.e. when
\beq
\label{sol1}
e_\Delta \le e_\mathrm{p} . 
\eeq

The relations between the intrinsic parameters of an oblate cluster and its observable features are \citep{ser+al06}
\begin{eqnarray}
 e_\mathrm{p}       & = &  \frac{1}{\sqrt{\left(1-q_{\mathrm{int}}^2\right) \cos ^2 i +q_{\mathrm{int}}^2}},  \label{obl1} \\
 e_\Delta  & = &  \sqrt{ 1 + \left(\frac{1}{q_{\mathrm{int}}^2} -1 \right) \cos ^2 i } . \label{obl2}
\end{eqnarray}
Then,
\begin{eqnarray}
 q_\mathrm{int}   & = &  \frac{1}{e_\mathrm{p} e_{\Delta }} ,  \label{obl3} \\
 \cos i           & = &  \sqrt{\frac{e_{\Delta }^2-1}{e_\mathrm{p}^2 e_{\Delta }^2-1}} . \label{obl4}
\end{eqnarray}
An oblate-like solution is admissible only when the size along the line of sight is larger than the maximum size in the plane of the sky, i.e. when
\beq
\label{sol2}
e_\Delta \ge 1 .
\eeq
Both the prolate and the oblate solutions are admissible at the same time only when
\beq
\label{sol3}
1 \le e_\Delta \le e_\mathrm{p} ,
\eeq
i.e. when the size along the line of sight is intermediate with respect to the projected dimensions. Even the assumption of axial symmetry is not enough in general to have an unique solution, but one have to specify if the cluster is assumed to be either prolate or oblate.

Let us now consider if, given a random orientation of the line of sight, it is more likely that the observed elongation and ellipticity are compatible with either a prolate or an oblate solution or with both of them. If clusters are randomly oriented, then they uniformly occupy the $\cos \theta_\mathrm{Eu}$ - $\varphi_\mathrm{Eu}$ plane. A cluster will be more likely interpreted as prolate than as oblate if the total area of the loci in the $\cos \theta_\mathrm{Eu}$-$\varphi_\mathrm{Eu}$ plane where only the prolate solution is admissible (grey regions in Fig.~\ref{Fig_ProOrObl}) is larger than that corresponding to the oblate case (black regions). In Fig.~\ref{Fig_ProOrObl}, we show the regions in the $\cos \theta_\mathrm{Eu}$-$\varphi_\mathrm{Eu}$ plane where a prolate and/or oblate solution is possible for three different sets of intrinsic axial ratio. You can also note the recurrence properties in the $\cos \theta_\mathrm{Eu}$-$\varphi_\mathrm{Eu}$. We consider a nearly prolate cluster with $e_1=1.3$, $e_2 =1.4$ ($T=0.75$), a nearly oblate ellipsoid with $e_1=0.7$, $e_2 =0.6$ ($T=0.25$) and a pure triaxial specimen, $e_1=1.2$, $e_2 =0.8$ ($T=0.5$). Even when the cluster is intrinsically close to either a prolate ($T=0.75$) or oblate ($T=0.25$) geometry, projection effects nearly completely hide this property and conflicting domains in the $\cos \theta_\mathrm{Eu}$ - $\varphi_\mathrm{Eu}$ plane have very similar extensions. However, for most of the orientations, both solutions are compatible with the observed quantities. These relative proportions change when the cluster get more triaxial ($T=0.5$). In that case the areas of the three different domains are similar.

In Fig.~\ref{Fig_ProOrObl_e1e2} we consider the most likely domain as a function of the intrinsic axial ratios. The most likely condition, for fixed values of $e_1$ and $e_2$, is determined by checking which existence conditions is fulfilled by elongation and projected ellipticity when averaged over the $\cos \theta_\mathrm{Eu}$ - $\varphi_\mathrm{Eu}$ plane, i.e. $\langle e_\Delta\rangle_{\varphi_\mathrm{Eu},\theta_\mathrm{Eu}} > \langle e_\mathrm{p} \rangle_{\varphi_\mathrm{Eu},\theta_\mathrm{Eu}}$ for a most likely oblate solution, $\langle e_\Delta\rangle_{\varphi_\mathrm{Eu},\theta_\mathrm{Eu}} < 1$ for a most likely prolate solution and $1 \le \langle e_\Delta\rangle_{\varphi_\mathrm{Eu},\theta_\mathrm{Eu}} \le \langle e_\mathrm{p} \rangle_{\varphi_\mathrm{Eu},\theta_\mathrm{Eu}}$ for a most likely double solution. For a large range of axial ratios, the cluster projections are compatible with both solutions. This holds in particular for the large central squared area in Fig.~\ref{Fig_ProOrObl_e1e2}, corresponding to $0.4 \ls e_1, e_2 \ls 2.5$, and in the region corresponding to nearly prolate clusters ($e_1 \sim e_2 >1$). We see that in the parameter range explored in Fig.~\ref{Fig_ProOrObl_e1e2}, $0.1 \le e_1, e_2 \le 10$, there are no locations where the case of only prolate solution is the most likely.

\section{Approximate deprojections}
\label{sec_appr}

\begin{figure*}
\begin{tabular}{c}
\resizebox{\hsize}{!}{\includegraphics{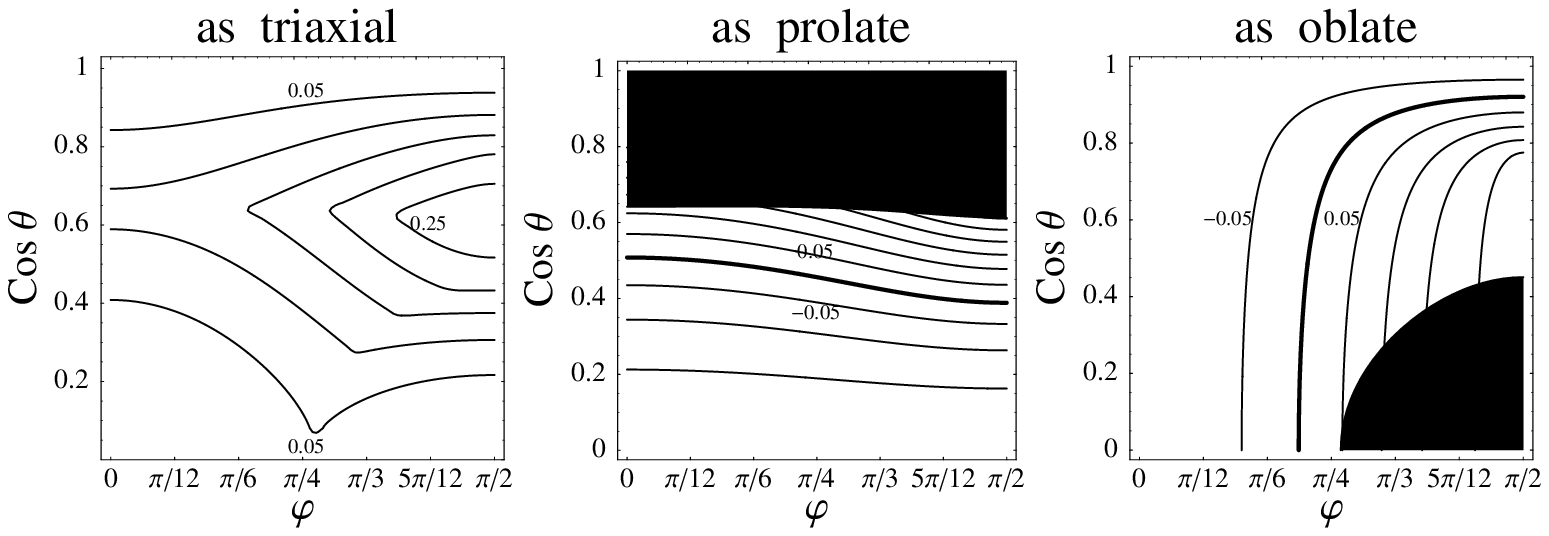}} \\
\resizebox{\hsize}{!}{\includegraphics{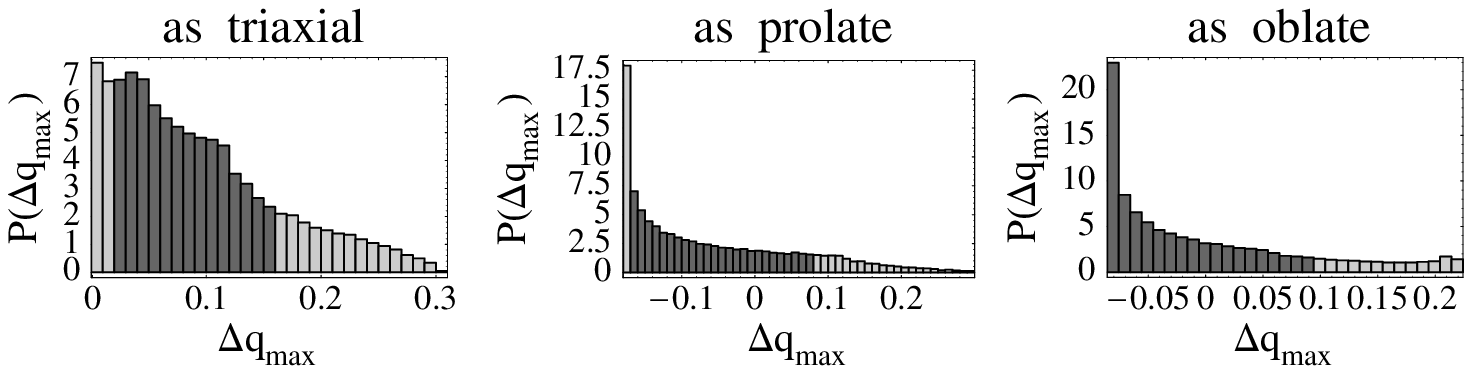}}
\end{tabular}
\caption{Deviation of the inferred maximum axial ratio from the actual one ($\Delta q_\mathrm{max}$). The intrinsic parameters are fixed to $e_1 =0.6$, $e_2 =0.7$ and $e_3=1$. $q_\mathrm{max}$ has been derived under the hypothesis of a triaxial ellipsoid aligned along the line of sight, of a prolate and of an oblate ellipsoid in the left, central and right panels, respectively. In the upper panels, $\Delta q_\mathrm{max}$ is plotted as a function of the Euler's angle of the line of sight. Contours of equal derived $q_\mathrm{max}$ are drawn in steps of $0.05$. The thicks line tracess the loci of points where the deprojection gives the actual $q_\mathrm{max}$. The hypotheses breaks down in the black regions. In the bottom panels, the normalised probability distribution of the deviation $\Delta q_\mathrm{max}$ is considered. Data are binned at intervals of $0.01$. The shadowed and light shadowed regions contain the $68.3\%$ and the whole range of the inferred values, respectively.}
	\label{Fig_qMax_ThetaPhi_025}
\end{figure*}

\begin{figure*}
\begin{tabular}{c}
\resizebox{\hsize}{!}{\includegraphics{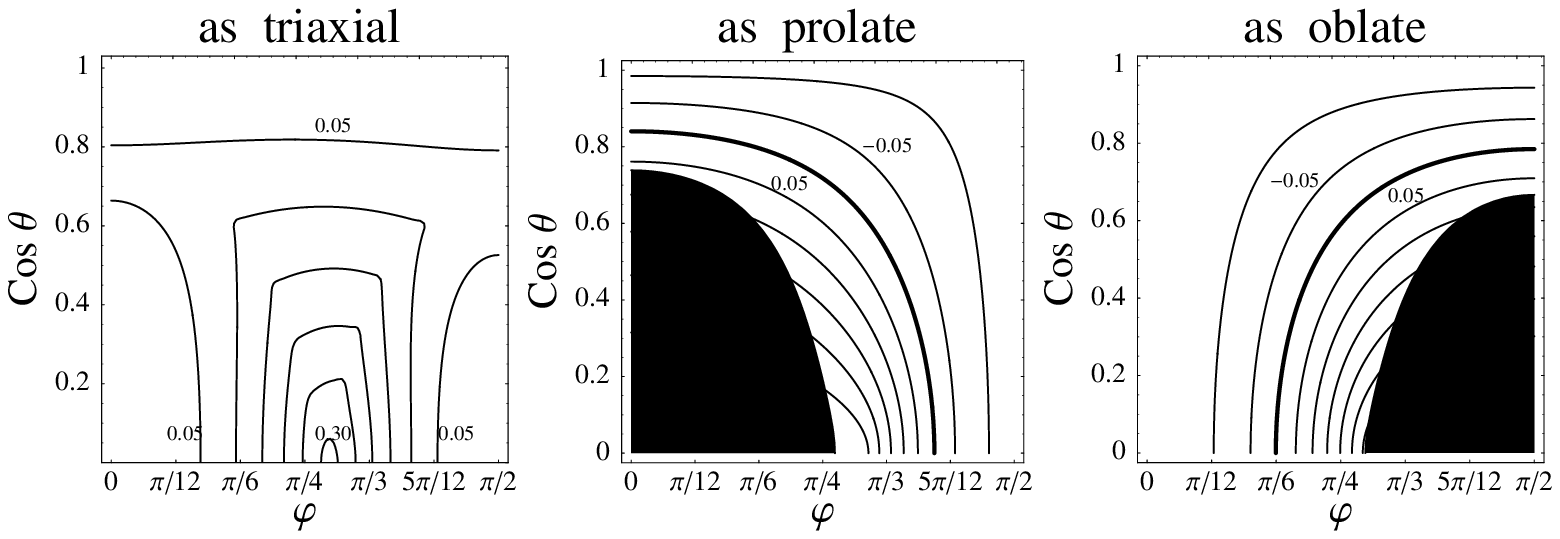}} \\
\resizebox{\hsize}{!}{\includegraphics{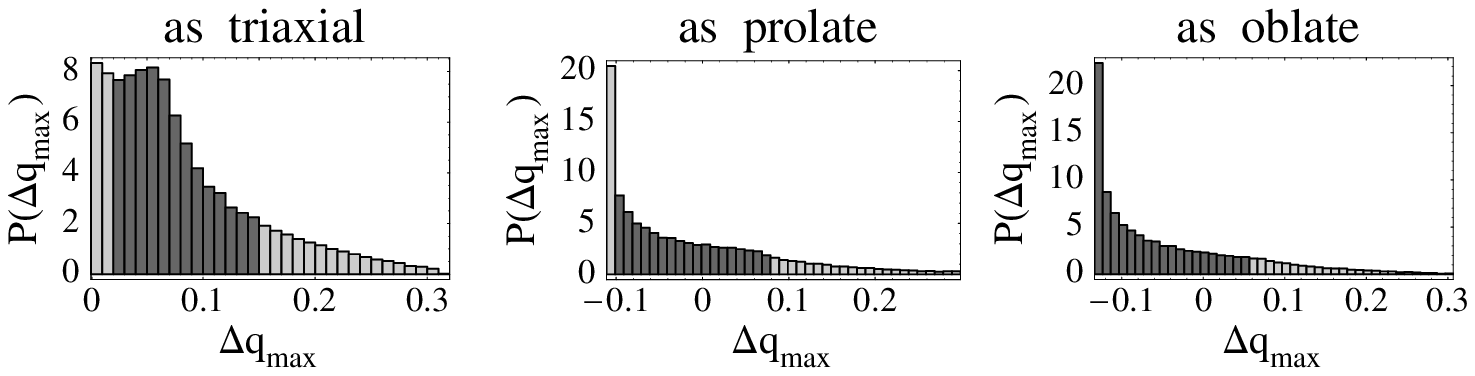}}
\end{tabular}
\caption{The same as in Fig.~\ref{Fig_qMax_ThetaPhi_025} for $e_1 =1.2$, $e_2 =0.8$ and $e_3=1$ ($T=0.5$).}
	\label{Fig_qMax_ThetaPhi_050}
\end{figure*}

\begin{figure*}
        \resizebox{\hsize}{!}{\includegraphics{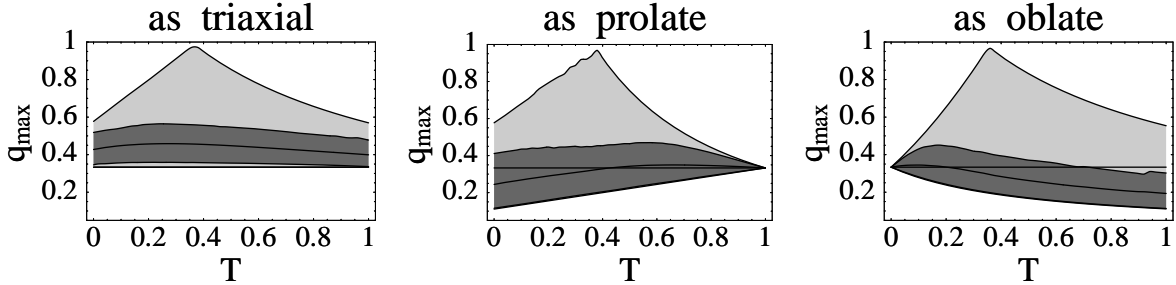}}
      \caption{Inferred axial ratio as a function of the triaxiality parameter $T$ under different hypotheses. The true $q_\mathrm{max}$ is fixed to $1/3$, as shown by the horizontal lines. The full line represents the inferred $q_\mathrm{max}$ averaged over the orientations of the line of sight, assumed to be randomly distributed. The shadowed and light shadowed regions contain the $68.3\%$ and the whole range of the inferred values, respectively. $q_\mathrm{max}$ has been derived under the hypothesis of a triaxial ellipsoid aligned along the line of sight, of a prolate and of an oblate ellipsoid in the left, central and right panel, respectively.}
	\label{Fig_qMax_T_033}
\end{figure*}

\begin{figure*}
        \resizebox{\hsize}{!}{\includegraphics{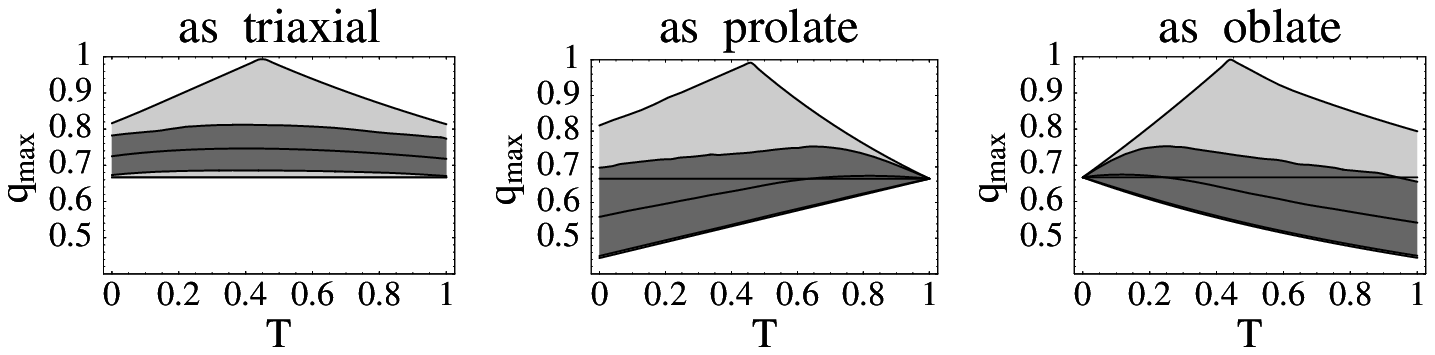}}
        \caption{The same as in Fig.~\ref{Fig_qMax_T_033} for an intrinsic $q_\mathrm{max} =2/3$.}
	\label{Fig_qMax_T_067}
\end{figure*}

Let us now consider the error made when the intrinsic shape of the cluster if inferred under the assumption that the cluster is either aligned with the line of sight or is axially symmetric. Since we are not interested in observational noise but just in the intrinsic error due to degeneracies in the deprojection, we suppose to have fiducial X-ray, SZE and GL data and that we can analyse them using the proper density distributions. This implies that we can get a correct measurements of both elongation and ellipticity from observations. Elongation and ellipticity depend only on the intrinsic ellipticity parameters, $e_1$, $e_2$ and $e_3$ and on the orientation of the line of sight, $\varphi_\mathrm{Eu}$ and $\theta_\mathrm{Eu}$, see Sec.~\ref{sec_proj}. As discussed in Sec.~\ref{sec_part}, $e_\Delta$ and $e_\mathrm{p}$ are all we need to perform the deprojection under the hypotheses of either a triaxial system aligned with the line of sight or an ellipsoid of revolution. Then, in order to evaluate the error made in the deprojection we have just to compare the intrinsic values (that we can fix at the beginning) to the ones inferred under the procedures outlined on Sec.~\ref{sec_part}. To this aim, it is useful to consider the maximum axial ratio, $q_\mathrm{max} = e_\mathrm{min}/e_\mathrm{max}$, i.e. the ratio of the minor to the major axis of the intrinsic ellipsoid. 

In Figs.~\ref{Fig_qMax_ThetaPhi_025} and~\ref{Fig_qMax_ThetaPhi_050} (upper panels), we consider the error made as a function of the line of sight orientation for two different sets of axial ratios: a nearly oblate case with $e_1=0.6$, $e_2 =0.7$ ($T=0.25$, $q_\mathrm{max} = 0.6$) in Fig~\ref{Fig_qMax_ThetaPhi_025} and a triaxial case with $e_1=1.2$, $e_2 =0.8$ ($T=0.5$, $q_\mathrm{max} = 2/3$) in Fig~\ref{Fig_qMax_ThetaPhi_050}. The inversion under the respective hypothesis can not be performed for the directions filling the black regions. The situation for a nearly prolate cluster with $T=0.75$ are very similar to the nearly oblate case with $T=0.25$ if we interchange the positions of the middle and of the right panel in Fig.~\ref{Fig_qMax_ThetaPhi_025}. Under the hypothesis that the line of sight is along one of the principal axes of the ellipsoid, the axial ratio will be over-estimated, i.e. the cluster will end up looking rounder. Obviously, the error is minimum when the lining up is pretty much satisfied ($\theta_\mathrm{Eu} \sim \pi/2$ and $\varphi_\mathrm{Eu} \sim 0$ or $\pi/2$, $\theta_\mathrm{Eu} \sim 0$). On the other hand, the loci of orientations where the deviation is maximum depend on the intrinsic geometry of the ellipsoid. For axial ratios close to the oblate case, see Fig~\ref{Fig_qMax_ThetaPhi_025}, or close to the prolate geometry ($T=0.75$), we have $\Delta q_\mathrm{max} \gs 0.25$ for $\varphi_\mathrm{Eu} \sim \pi/2$ and $\cos \theta_\mathrm{Eu} \sim 0.6$. For a $T=0.5$ shape, see Fig~\ref{Fig_qMax_ThetaPhi_050}, the deviation is maximum for $\varphi_\mathrm{Eu} \sim \pi/4$ and $\theta_\mathrm{Eu} \ls \pi/2$, where $\Delta q_\mathrm{max} \gs 0.30$.

Under the hypothesis of axial symmetry, the maximum axial ratio can be either over- or under-estimated. Let us first consider the nearly oblate intrinsic shape represented in Fig.~\ref{Fig_qMax_ThetaPhi_025}. Under the assumption of oblate geometry (right panel), for most of the orientations the error will be in the range $ -0.1 \ls \Delta q_\mathrm{max} \ls 0.1$; the maximum deviation, $\Delta q_\mathrm{max} \ls 0.25$, occurs near the forbidden region for large values of $\varphi_\mathrm{Eu}$ ($\sim \pi/2$). Deviations will be much larger under the assumption of a prolate geometry (middle panel). It can be easily seen that the region where $-0.1 \ls \Delta q_\mathrm{max} \ls 0.1$ is much smaller than the corresponding one obtained under the oblateness assumption. For a triaxial intrinsic shape, Fig.~\ref{Fig_qMax_ThetaPhi_050}, the oblate and the prolate hypotheses works pretty much in the same way.

The normalised probability distributions of the deviation $\Delta q_\mathrm{max}$ are plotted in the bottom panels of Figs.~\ref{Fig_qMax_ThetaPhi_025} and~\ref{Fig_qMax_ThetaPhi_050}. The distributions have been obtained under the hypothesis of randomly oriented clusters, so that the probability of a deviation, $P(\Delta q_\mathrm{max}) d\Delta q_\mathrm{max}$ is proportional to the area in the $\varphi_\mathrm{Eu}$-$\cos \theta_\mathrm{Eu}$ plane of the region where the values of the deviation are between $\Delta q_\mathrm{max}$ and $ \Delta q_\mathrm{max} + d\Delta q_\mathrm{max}$. The portions of the grey regions above or below the averaged value refer to the same total area in the $\cos \theta_\mathrm{Eu}$-$\varphi_\mathrm{Eu}$ plane. Both the light grey regions below and above the average correspond to a region in the $\cos \theta_\mathrm{Eu}$-$\varphi_\mathrm{Eu}$ plane whose area is 15.8\% of the total. The fact that the light grey region below the average has a smaller extension than that above means that in the corresponding region in the $\cos \theta_\mathrm{Eu}$-$\varphi_\mathrm{Eu}$ plane the range of $\Delta q_\mathrm{max}$ values is much smaller than in the other one. As general feaures, we see that the aligned triaxial hypothesis over-estimate the axis ratio and that, in general, there is always a long tail for large positive values of $\Delta q_\mathrm{max}$, i.e. there is always a chance that the cluster appears nearly round. If the case of $T=0.25$, the average value of the deviation is $0.09$, $-0.06$ and $\ls 0$ for an aligned triaxial, prolate or oblate assumption, respectively.

The total range of inferred $q_\mathrm{max}$ under different hypotheses and for various orientations of the line of sight depends in general on the triaxiality degree of the cluster. In Figs.~\ref{Fig_qMax_T_033} and~\ref{Fig_qMax_T_067}, we show the general features of the inferred axial ratio as a function of the triaxiality parameter $T$ for two different intrinsic axial ratios, $q_\mathrm{max} =1/3$ and $2/3$, respectively. In particular, we consider the value of the inferred $q_\mathrm{max}$ averaged over all the possible orientations of the line of sight, the range including the 68.3\% of the inferred axial ratios around the averaged value (grey regions), and the total range of inferred values (light shadowed region). These values try to summerize the full information contained in the $\theta_\mathrm{Eu}$-$\varphi_\mathrm{Eu}$ plane, as shown for single $T$ values in Figs.~\ref{Fig_qMax_ThetaPhi_025} and~\ref{Fig_qMax_ThetaPhi_050}. 

The extension of the grey regions in Figs.~\ref{Fig_qMax_T_033} and~\ref{Fig_qMax_T_067} has the same meaning as in the bottom panels of  Figs.~\ref{Fig_qMax_ThetaPhi_025} and~\ref{Fig_qMax_ThetaPhi_050}. If we  make the hypothesis of a triaxial cluster aligned with the line of sight, the error is nearly independent of the effective triaxiality. The axial ratio $q_\mathrm{max}$ will be overestimated in average by $0.07 (0.05) \ls \Delta q_\mathrm{max} \ls 0.12 (0.08)$ for an actual intrinsic axial ratio of $e_\mathrm{min}/e_\mathrm{max}=1/3$ $(2/3)$. The range for the $68.3\%$ of the orientations is of $0.01 (0.01) \ls \Delta q_\mathrm{max} \ls 0.23 (0.14)$, whereas the total range spans from the true value to a maximum of $q_\mathrm{max} \sim 1$, i.e an apparently round geometry, for $T \sim 0.4$ ($0.45$). 

If we make the hypothesis that the cluster is prolate, the error decreases with the cluster effectively approaching a nearly prolate shape ($T \rightarrow 1$). For $e_\mathrm{min}/e_\mathrm{max}=1/3$ $(2/3)$, the average deviation is as large as $\gs -0.09$ ($-0.11$) for $T \rightarrow 0$, and is really small ($|\Delta q_\mathrm{max}| \ls 0.02$) for $T \gs 0.4 (0.5)$. The range for the $68.3\%$ of the orientations is of $-0.22 (-0.22) \ls \Delta q_\mathrm{max} \ls 0.08 (0.03)$ for $T=0$, $-0.13 (-0.08) \ls \Delta q_\mathrm{max} \ls 0.12 (0.09)$ for $T \simeq 0.4$ ($0.6$) and goes to $0$ for $T \rightarrow 1$. The full range is of $-0.22 (-0.22) \ls \Delta q_\mathrm{max} \ls 0.24 (0.15)$ for $T=0$, $-0.13 (-0.11) \ls \Delta q_\mathrm{max} \ls 0.59 (0.29)$ for $T \simeq 0.4$ $(0.5)$  and, as usual, goes to $0$ for $T \rightarrow 1$. 

Under the hypothesis of an oblate geometry the situation is reversed. For $e_\mathrm{min}/e_\mathrm{max}=1/3$ $(2/3)$, the average deviation is small ($|\Delta q_\mathrm{max}| \ls 0.02$) for $T \ls 0.3$ $(0.4)$ and goes to $\sim -0.14$ $(-0.13)$ for $T \rightarrow 1$. The range for the $68.3\%$ of the orientations starts from $0$ at $T=0$, is of  $-0.09 (-0.06) \ls \Delta q_\mathrm{max} \ls 0.11 (0.08)$ for $T \simeq 0.2$ ($0.2$) and goes to $-0.22 (-0.22) \ls \Delta q_\mathrm{max} \ls -0.03 (-0.01)$ for $T=1$. The full range is of $ -0.14 (-0.12) \ls \Delta q_\mathrm{max} \ls 0.59 (0.32)$ for $T \simeq 0.4$ ($0.45.$) and  $-0.22 (-0.22)\ls \Delta q_\mathrm{max} \ls 0.22 (0.3)$ for $T=1$.

\section{Bad identifications of axially symmetric ellipsoids}
\label{sec_badl}

As seen in Sec.~\ref{sec_part}, even under the strong assumption that we are observing an ellipsoid of revolution, the derivation of the intrinsic parameters is in general not unique. In fact, for an axially symmetric cluster whose size along the line of sight is intermediate with respect to its width and the length in the plane of the sky, Eq.~(\ref{sol3}), we can not establish if it is either prolate or oblate.

\subsection{Prolate as oblate}

\begin{figure*}
        \resizebox{\hsize}{!}{\includegraphics{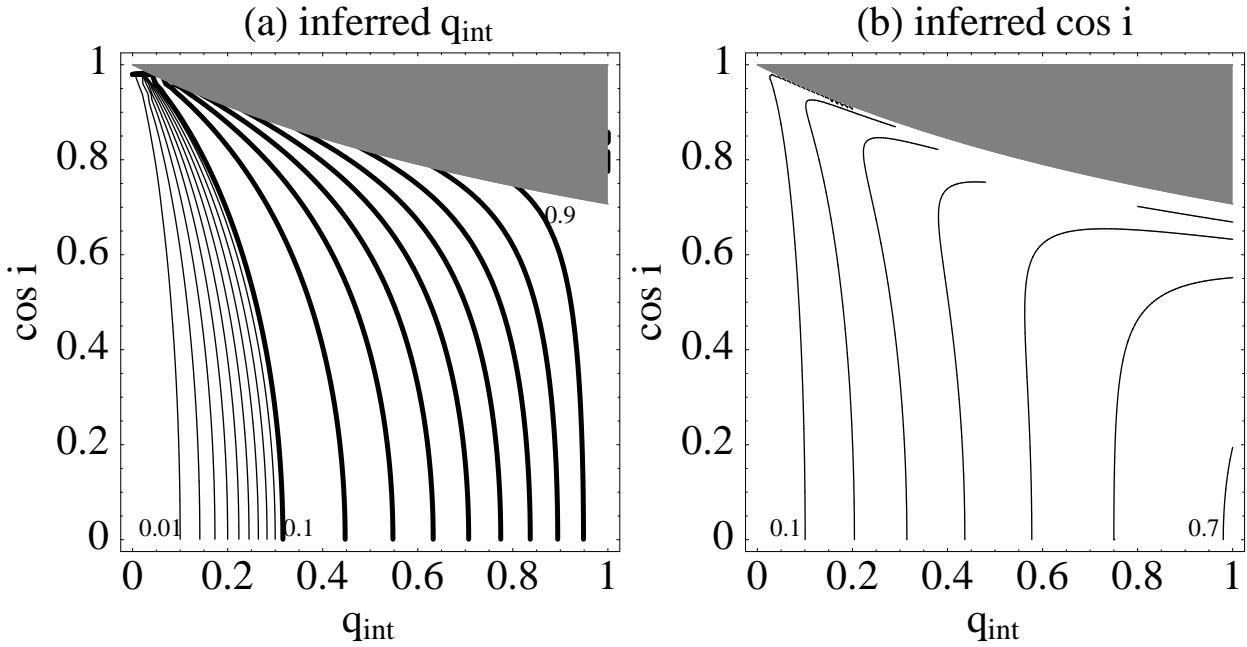}}
        \caption{Intrinsic (wrong) parameters of an intrinsically prolate ellipsoids inferred under the oblate hypothesis as a function of the true intrinsic parameters. In the shadow regions, the wrong inversion can not be performed. Left (a) panel: inferred axial ratio $q_\mathrm{int}$. Thick contour values run from $0.9$ to $0.1$ with steps of $0.1$; thin contour values run from $0.09$ to $0.01$ with steps of $0.01$. Right (b) panel: inferred inclination  $\cos i$. Contour values run from $0.1$ to $0.7$ with steps of $0.1$.
}
\label{Fig_ProAsObl}
\end{figure*}

A prolate cluster can have the same projected map of an oblate system. An oblate-like deprojection for a prolate ellipsoid is possible only when Eq.~(\ref{sol2}) holds. In terms of the intrinsic parameters of the prolate cluster, the inclination angle has to be greater than a given threshold,
\beq
\label{asObl1}
\cos i \le \frac{1}{\sqrt{1+q_\mathrm{int}}};
\eeq
if $i \ge \pi/4$, there is always a possible, but wrong, oblate-like solution. As can be seen from  Fig.~\ref{Fig_ProAsObl}, assuming a population of prolate ellipsoids randomly oriented, most of them will project in the sky in a way that is compatible with an oblate morphology.

An erroneous hypothesis on the morphology affects the estimate of the intrinsic parameters. Here, let us consider the error made when considering as oblate an intrinsically prolate cluster. If Eq.~(\ref{asObl1}) holds, the inversion can be performed to infer wrong values for $q_\mathrm{int}$ and $i$ which are still compatible with the observations. To find such wrong parameters and infer the error made badly identifying  a prolate system for an oblate one, we have to substitute in Eqs.~(\ref{obl3},~\ref{obl4}) the expressions of $e_\mathrm{p}$ and $e_\Delta$ for a prolate cluster, Eqs.~(\ref{pro1},~\ref{pro2}). The inferred axial ratio and cosine of the inclination angle are
\beq
\frac{q_{\mathrm{int}}^2}{\left[ 1 - \left(1 -q_{\mathrm{int}}^2 \right) \cos ^2 i \right]^{3/2}}, \label{asObl2} 
\eeq
and
\beq
q_{\mathrm{int}} \sqrt{\frac{\left[ 1 - \left(1- q_{\mathrm{int}}^2 \right) \cos^2 i  \right]^2-q_{\mathrm{int}}^2}{\left[  1 -  \left(1 - q_{\mathrm{int}}^2 \right) \cos ^2 i \right]^3-q_{\mathrm{int}}^4}}   \label{asObl3}
\eeq
respectively. Results are shown in Fig.~\ref{Fig_ProAsObl}. In general, the inferred axial ratio is less than the actual one. The inferred wrong parameters are sensitive to the intrinsic axial ratio but depend very weakly on the inclination angle. In general the inferred axial ratio increases with the true axial ratio. From Eqs.~(\ref{asObl1},~\ref{asObl2}), we get an upper limit for the inferred wrong axial ratio, which is $\le \sqrt{q_\mathrm{int}}$.
For $q_\mathrm{int} \ls 0.4$, the inferred value trails the actual one by 0.2; for $q_\mathrm{int} \gs 0.7$, the gap is reduced to $\sim 0.1$. Whatever $i$, an intrinsically pretty elongated prolate cluster ($q_\mathrm{int} \ls 0.2$) can be misidentified for an oblate object with a large inclination ($\cos i \ls 0.2$). On the other hand, a nearly round even if slightly prolate object ($q_\mathrm{int} \gs 0.8$), can appear as an oblate cluster with $0.6 \ls \cos i \ls 0.7$.

\subsection{Oblate as Prolate}

\begin{figure*}
        \resizebox{\hsize}{!}{\includegraphics{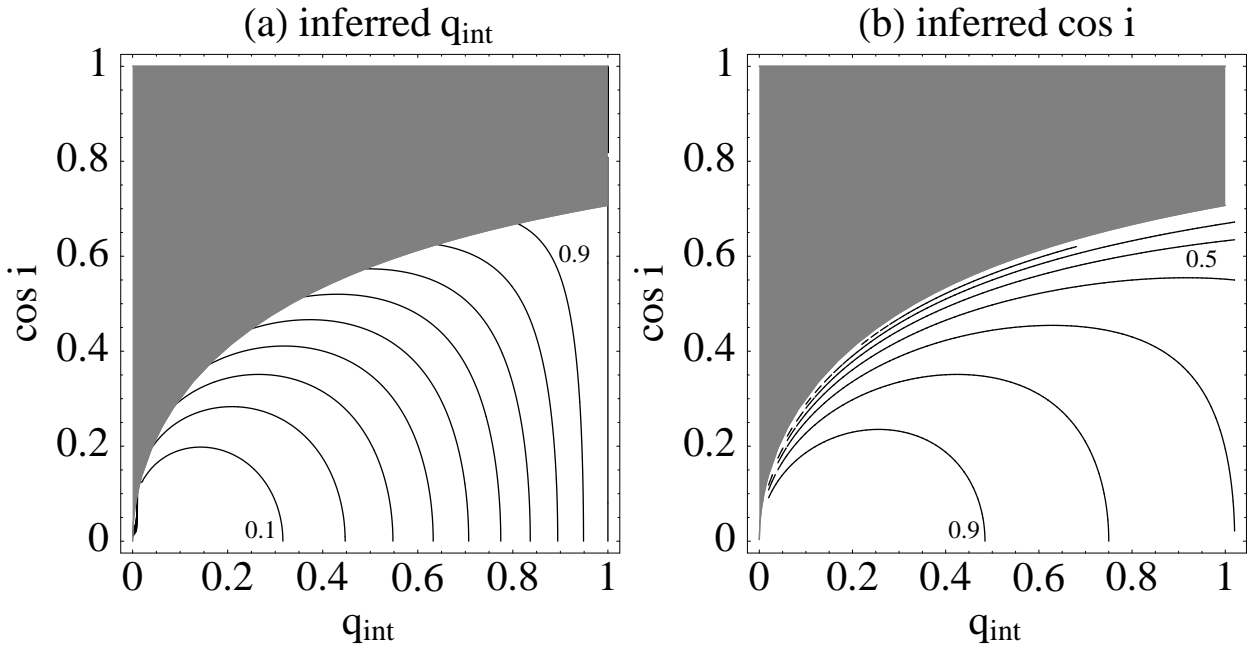}}
        \caption{Intrinsic (wrong) parameters of an intrinsically oblate ellipsoid inferred under the prolate hypothesis as a function of the true intrinsic parameters. In the shadowed regions, the wrong inversion can not be performed. Left (a) panel: inferred axial ratio $q_\mathrm{int}$. Contour values run from $0.1$ to $1$ with steps of $0.1$. Right (b) panel: inferred inclination  $\cos i$. Contour values run from $0.9$ to $0$ with steps of $0.1$. For inferred values smaller than $0.4$, contours are pretty close to each other.
}
        \label{Fig_OblAsPro}
\end{figure*}

Let us now consider when and how an oblate cluster is projected in a way compatible with a prolate geometry. As can be seen from Eq.~(\ref{sol1}) and Eqs.~(\ref{obl1},~\ref{obl2}), the condition to be fulfilled is
\beq
\label{asPro1}
\cos i \le \sqrt{ \frac{q_\mathrm{int}}{1+q_\mathrm{int} } }.
\eeq
Comparing the areas in Fig.~\ref{Fig_ProAsObl} and~\ref{Fig_OblAsPro} where a double solution is allowed and assuming a random orientation, we can see how is most likely that a prolate objects is interpreted as oblate that the inverse case. Under the assumption that an oblate cluster is prolate, one will get the wrong intrinsic axial ratio
\beq
 \left[ 1+ \left(\frac{1}{q_{\mathrm{int}}^2}-1\right) \cos ^2 i \right]^{3/2} q_{\mathrm{int}}^2, \label{asPro2}  
\eeq
and the wrong inclination (expressed in terms of the cosine)
\beq
\sqrt{\frac{q_{\mathrm{int}}^2-\left[  \left(1-q_{\mathrm{int}}^2\right) \cos^2 i +q_{\mathrm{int}}^2\right]^2}{q_{\mathrm{int}}^2-\left[  \left(1-q_{\mathrm{int}}^2\right) \cos ^2 i +q_{\mathrm{int}}^2   \right]^3}} \label{asPro3} .
\eeq
As can be seen from Eqs.~(\ref{asPro1},~\ref{asPro2}), the upper limit for the inferred wrong axial ratio is $\sqrt{q_\mathrm{int}}$, as in the previous case. Results are shown in Fig.~\ref{Fig_OblAsPro}. The inferred axial ratio is less than the actual one, i.e. to compensate for the wrong assumption, the cluster is supposed to be more elliptical. Errors can be very large. An highly flattened oblate object nearly edge-on ($q_\mathrm{int} \ls 0.2, \cos i \ls 0.2$) can be interpreted as a very elongated prolate cluster nearly face-on.

\section{Discussion}
\label{sec_disc}

The combined analysis of X-ray, SZE and GL maps can determine the elongation of the cluster along the line of sight, its width and length in the plane of the sky and an unbiased estimate of the Hubble constant. Even if the temperature, metallicity and the density profile of the ICM can be determined, the shape and orientation of the cluster can not be fully constrained.

The restrictive assumption of an axially symmetric geometry is in general not enough. In fact, a prolate cluster can cast on the plane of the sky in the same way of an oblate ellipsoid with different inclination and axial ratio and vice versa. Then, in general one must assume that the cluster is either prolate or oblate. Even if for some particular combinations of orientation and intrinsic ellipticity only one axially symmetric solution is possible, there are always some more triaxial configurations compatible with data.

Under the hypothesis that the cluster is prolate or oblate, the inversion is likely to estimate the cluster more elliptical than the actual value. The main shortcoming is that forcing an axially symmetric geometry to fit an actually triaxial cluster can strongly bias the analysis: even the projection of a nearly prolate, but intrinsically triaxial cluster can be compatible, for some orientations of the system, only with an oblate geometry but not compatible with a prolate one and vice-versa. On the other hand, assuming that the cluster is triaxial and aligned with the line of sight is a more conservative approach. The cluster looks rounder and the minor to major axis ratio is likely to be overestimated by $\sim 0.1$, but we are assured that the intrinsic geometry is not completely mistaken.

A very useful additional information for the inversion would be the knowledge of the orientation in the plane of the sky of the projection of one of the intrinsic axes of the ellipsoid. Whereas this can not be obtained by analysing projected maps, other approaches could be viable. Unless clusters are axially symmetric bodies with internal streaming motions about a fixed symmetry axis, line of sight rotational motions along apparent major axis are in general associated with velocity gradients along apparent minor axes \citep{con56}. This effect can then be studied under suitable approximations, i.e. that the velocity field is made up of an overall figure rotation about either the shortest or the longest principal axis together with internal streaming around that axis \citep{bin85}. These theoretical velocity fields can then be projected on to the plane of the sky and compared with observations. However, current detections of velocity gradients in galaxy clusters are still uncertain \citep[and references therein]{hw+le07,ser07}.

A more direct approach to map out the three dimensional structure of a rich cluster is through distances measurements for individual members. Distances could be obtained using the method of surface brightness fluctuations \citep{mei+al07} or empirical relationships based on Tully-Fisher or $D$-$\sigma$ distance indicators \citep{mas+al06}. This method is limited to very nearby clusters, but would give a complete 3-D description of the cluster galaxy distribution. 

The theoretical analysis performed in this paper has the advantage of clearly showing the main degeneracies of the inversion procedure. The simple isothermal $\beta$ model was enough to show what we can learn on the 3-D structure of galaxy clusters from projected maps. Such an analysis avoided some final $\chi^2$ fitting procedure to some particular simulated map, a method which could miss some important facets of the degeneracy question.

In this paper, I have considered projected X-ray, SZE and GL maps. In any case, luminosity and surface brightness observations in the optical band would provide the same kind of information. Furthermore, if some assumptions are relaxed and we let the galaxy distribution have a different intrinsic shape with respect to the ICM, we should determine two more intrinsic axial ratios. In any case, some other observational constraints could be used in this case. Assuming that galaxy and ICM 3-D distributions have the same orientation but different ellipticities, then their isophotes would be misaligned in projection \citep{ro+ko98}.

The hydrostatic equilibrium assumption, used in this paper to relate GL observations to the other maps, could be substituted by other theoretical hypotheses. The assumption of a measurable and constant baryon fraction has been suggested as a different way to break the degeneracy between the Hubble constant and the elongation \citep{coo98}. However such different theoretical assumptions all aim to break the same degeneracy. If used together they would over-constrain some facets of the problem, but would not shed light on other undetermined features, such as the orientation issue.

\section*{Acknowledgements}
I thank E. De Filippis for the many discussions on the topic. M.S. is supported by the Swiss National Science Foundation and by the Tomalla Foundation.

\bibliographystyle{mn2e}

\setlength{\bibhang}{2.0em}

\end{document}